\documentclass[sigconf]{acmart}
\let\labelindent\relax
\usepackage{enumitem}

\setlist[enumerate]{wide=\parindent}

\usepackage{lipsum}

\usepackage{lmodern}
\usepackage{amssymb,amsmath}
\usepackage{ifxetex,ifluatex}
\usepackage{fixltx2e} 
\ifnum 0\ifxetex 1\fi\ifluatex 1\fi=0 
  \usepackage[T1]{fontenc}
  \usepackage[utf8]{inputenc}
\else 
  \defaultfontfeatures{Ligatures=TeX,Scale=MatchLowercase}
\fi
\IfFileExists{upquote.sty}{\usepackage{upquote}}{}
\IfFileExists{microtype.sty}{%
\usepackage{microtype}
\UseMicrotypeSet[protrusion]{basicmath} 
}{}
\hypersetup{
            pdftitle={MXNET-MPI: Embedding MPI parallelism in Parameter Server Task Model for scaling Deep Learning},
            pdfauthor={; Amith R Mamidala},
            pdfborder={0 0 0},
            breaklinks=true}
\usepackage{url}

\usepackage{breakurl}
\usepackage{listings}
\usepackage{graphicx,grffile}
\makeatletter
\def\maxwidth{\ifdim\Gin@nat@width>\linewidth\linewidth\else\Gin@nat@width\fi}
\def\maxheight{\ifdim\Gin@nat@height>\textheight\textheight\else\Gin@nat@height\fi}
\makeatother
\setkeys{Gin}{width=\maxwidth,height=\maxheight,keepaspectratio}
\IfFileExists{parskip.sty}{%
\usepackage{parskip}
}{
\setlength{\parindent}{0pt}
\setlength{\parskip}{6pt plus 2pt minus 1pt}
}
\ifx\paragraph\undefined\else
\let\oldparagraph\paragraph
\renewcommand{\paragraph}[1]{\oldparagraph{#1}\mbox{}}
\fi
\ifx\subparagraph\undefined\else
\let\oldsubparagraph\subparagraph
\renewcommand{\subparagraph}[1]{\oldsubparagraph{#1}\mbox{}}
\fi

\let\origthelstnumber\thelstnumber
\makeatletter
\newcommand*\Suppressnumber{%
  \lst@AddToHook{OnNewLine}{%
    \let\thelstnumber\relax%
    \advance\c@lstnumber-\@ne\relax%
  }%
}
\newcommand*\Reactivatenumber{%
  \lst@AddToHook{OnNewLine}{%
    \let\thelstnumber\origthelstnumber%
    \advance\c@lstnumber\@ne\relax}%
}
\begin{document}

\title{MXNET-MPI: Embedding MPI parallelism in Parameter Server Task Model for scaling Deep Learning}
\author{Amith R Mamidala, Georgios Kollias, Fausto Artico}
 \affiliation{IBM T J Watson Research Center\\
 Yorktown Heights, New York, USA}
\email{amithr,gkollias,fausto.artico@us.ibm.com}
\author{Chris Ward}
\affiliation{IBM, Hursley Park\\
Hursley, UK}
\email{tjcw@us.ibm.com}



\begin{abstract}
Existing Deep Learning frameworks exclusively use either Parameter Server(PS) approach or MPI parallelism. In this paper, we discuss the drawbacks of such approaches and propose a generic  framework supporting both PS and MPI programming paradigms, co-existing at the same time. The key advantage of the new model is to embed the scaling benefits of MPI parallelism into the loosely coupled PS task model. Apart from providing a practical usage model of MPI in cloud, such framework allows for novel communication avoiding algorithms that do parameter averaging in Stochastic Gradient Descent(SGD) approaches. We show how MPI and PS models can synergestically apply algorithms such as Elastic SGD to improve the rate of convergence against existing approaches. These new algorithms directly help scaling SGD clusterwide. Further, we also optimize the critical component of the framework, namely global aggregation or allreduce using a novel concept of tensor collectives. These treat a group of vectors on a node as a single object allowing for the existing single vector algorithms to be directly applicable. We back our claims with sufficient emperical evidence using large scale ImageNet 1K data. Our framework is built upon MXNET but the design is generic and can be adapted to other popular DL infrastructures. 

\end{abstract}
\maketitle

\section{Introduction}\label{sec:introduction}
As Deep Learning(DL) continues its dominance in a multitude of disciplines such as Image Classification, Speech Recognition and Natural Language Processing, the need for DL systems of scale to reduce training times gains utmost importance. With scientists exploring newer and scalable algorithms, innovation in DL infrastructure and frameworks is critical to realize their potential on massively large supercomputers. For example, clusters of GPUs interconnected by high performance networks are being deployed and a major emphasis is on cloud to lower costs~\cite{aws,azure,google-cloud,ibm-cloud}. Also, future generation machines such as Sierra and Summit\cite{coral} would deploy thousands of nodes featuring IBM Power9 processors with multiple NVIDIA Volta GPUs per node interconnected by fast InfiniBand networks. 

Almost all of the existing DL frameworks adopt either a Parameter Server (PS) approach or use MPI parallelism to scale DL algorithms. MXNET~\cite{mxnet}, TensorFlow~\cite{abadi2016tensorflow} use PS. CNTK~\cite{CNTK}, Caffee~\cite{caffeMPIindiana,s-caffe,dl-cots} use MPI parallelism. The core computation done in these algorithms is a parallel Stochastic Gradient Descent or SGD. In the loosely coupled task model of PS approach, parallel SGD faces issues such as network hot-spots and slow convergence due to parameter staleness at scale. However, it is ubiquitous in cloud computing as it is inherently fault tolerant and elastic. On the other hand, MPI has been proven to deliver performance at scale but lacks a good fault-tolerant support, though ULFM~\cite{ulfm} and its implementations~\cite{mpich-ulfm,openmpi-ulfm,ulfm-issues},are promising steps in this direction. Also, dynamic sizing of MPI jobs poses several constraints~\cite{mpi-dynamic-issues}.  Both represent opposite ends of the programming space and considerable efficiencies can be achieved by using both at the same time.

Further, as number of workers doing parallel SGD increase, the algorithm and the system imposes a restriction on how much scaling is permissible without degrading the performance of the algorithm. For example, one of the important parameter of the SGD is the mini-batch size which cannot be indefinitely increased due to drop in the accuracy~\cite{largebatch-degrades,bengio2012practical} and reduced parallel efficiencies due to increasing communication costs~\cite{dl-intel}. Hence, a very important and pertinent question to ask is are the existing DL frameworks suitable for cluster wide scaling?

A crucial step for performance in SGD is the aggregation of gradient or parameter vectors of the model from each GPU and across the nodes. With the architectures offering new CPU, GPU topologies interconnected by high bandwidth networks such as NVLINK\cite{coral} can the existing collective algorithms operating on one vector per worker still apply?

In this paper, we focus on these issues. In particular,

1) We design, implement, evaluate a MXNET based framework supporting both PS and MPI parallelism at the same time by embedding MPI based collective primitives in the computation graph. The central idea is to make an independent MPI\_COMM\_WORLD job client to the PS. The number of clients is tunable, offering knobs for a smooth transition from PS on one end to pure MPI on the other end. This generic design offers flexibility to work with a variety of algorithms across traditional HPC and cloud based infrastructures.

2) Using large scale ImageNet 1K~\cite{imagenet} we demonstrate the key benefits of our approach, namely better scaling by reducing network contention and alleviating parameter staleness. The new model allows this by reducing number of clients and instead scaling each client. Compared to the default PS approach, our method improves the time per epoch by six times and also improves the rate of convergence at the same time.


3) We design a new MPI Elastic SGD algorithm allowing synchronous SGD methods within a MPI communicator and asynchronous lazy update of parameters outside. This allows new opportunities for scaling DL algorithms cluster wide. On the ImageNet we show more than 2X improvement in rate of convergence compared to all the major approaches of parallel SGD.

4) We demonstrate a new class of multi-node tensor collectives.
The central idea used in these primitives is to design collective operations around a GPU tensor per worker instead of a single vector per GPU. Using our optimizations, we arrive at validation accuracies over 0.72 for the complete ImageNet 1K training on a IBM Minsky cluster~\cite{ibm-minsky}.

We now describe the SGD optimization method, the fundamental numerical gradient optimization used in almost all the popular frameworks.

\section{SGD}\label{sec:sgd}
In this section, we explain the major issues in parallel and distributed SGD. This motivates the need for appropriate MPI adaptations of these algorithms for scaling.
\subsection{Mini-Batch SGD}\label{sec:mini-batch-stoch}
In a mini-batch SGD, the entire data is divided into several mini-batches, collectively known as the ``epoch". The computation iterates over the epoch, one mini-batch at a time. The model parameters at iteration t, $w_{t}$ are updated by an increment $\Delta w$ to get the parameters for the next iteration.
\begin{equation}\label{eq:param-update}
w_{t + 1} = w_{t} + \Delta w
\end{equation}
$\Delta w$ is computed as $\Delta w  = - \eta g$,
where $g$ is the gradient, $\eta$ is a hyper parameter called
as the learning rate. For the deep learning models, the model parameters and gradients are associated with the different network links across the layers. The gradients are obtained after doing a forward pass and then an auto-differentiation in the backward step. The final gradient, $g$ is the average of all the gradients obtained from the data samples in a mini-batch. Also, since the gradients are obtained as soon as a backward step for a layer is computed, these can be aggregated in parallel with the backward phase of the previous layer for parallel SGD described below. 


\subsection{Parallel SGD}\label{sec:parallel-sgd}
In parallel SGD\cite{ruder2016overview,lian2015asynchronous,recht2011hogwild},  each worker processes an independent portion of the data set. The many versions of the parallel SGDs differ in the manner gradients and parameters are computed and how the mini-batches are constructed. In parallel synchronous SGD, the mini-batch is divided across all the workers and all workers wait for global average  of locally computed gradients, $g$ before computing the next set of parameters, $w_{t+1}$, equation~\ref{eq:param-update}. 
In asynchronous form,  each worker gets a separate mini-batch and doesn't aggregate gradients from other workers. After every iteration, it interacts with a PS only to push the locally computed averaged gradient, $g$ and pull the latest parameters,$w_{t + 1}$  for their next mini-batch~\cite{google-DNN}. Please note that we use PS interchangeably to mean the model and also refer to the parameter servers of the framework.\\
New algorithms such as Elastic Averaging\cite{zhang2015deep}, Federated Averaging~\cite{federated-averaging} allow further decoupling with the PS.  In stead of interacting with PS after every iteration, they compute the weights locally and use PS to lazily compute the average of weights $w_{t}$ rather than gradients for the next mini-batch. 
For elastic averaging, the PS stores an additional set of model weights called as center variables, $\tilde{w}$. The update~\ref{elastic1} is done on the server and ~\ref{elastic2} is done on the client. 
$\alpha$ is another hyper parameter, like the learning rate of SGD, passed to the
algorithm. 
\begin{equation}\label{elastic1}
\tilde{w}_{\text{t+1}} =
\tilde{w}_{\text{t}} + \alpha*\left(w_{\text{t}} -
  {\tilde{w}}_{\text{t}} \right) 
  \end{equation}
 \begin{equation}\label{elastic2}
w_{\text{t+1}} = 
w_{\text{t}} - \alpha*\left(w_{\text{t}} -
{\tilde{w}}_{\text{t}}\right)
\end{equation}

\subsection{Issues with Parallel SGD}\label{sec:issues}
\begin{figure*}[h]
    \includegraphics[width=\textwidth]{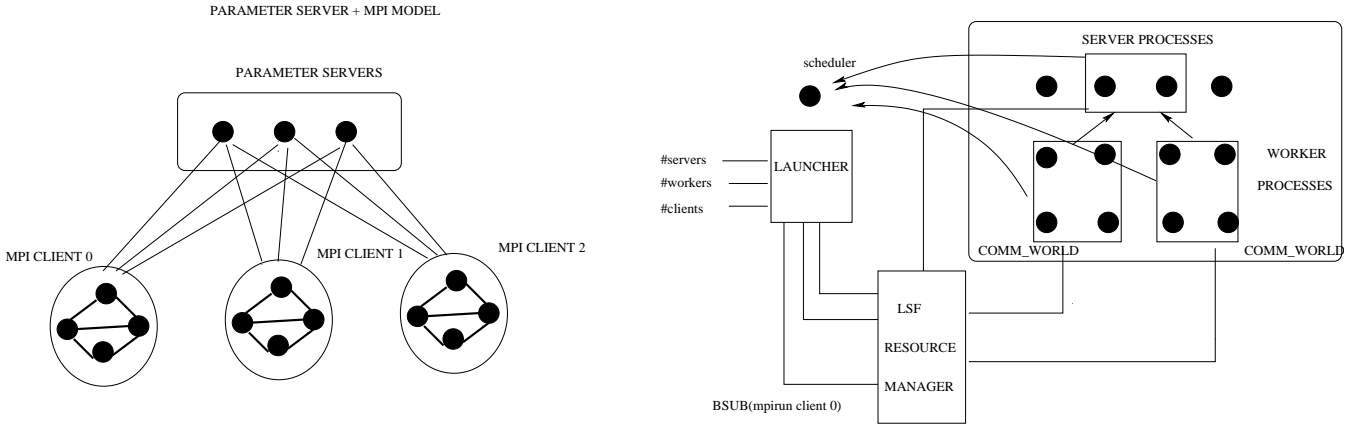}
    \vspace{-1mm}
  \caption{MPI + Parameter server}
  \label{fig:ps-mpi-model}
\end{figure*}

We now explain the different issues with parallel SGDs.\\
\textbf{Network Contention:} One of the major issues faced with Synchronous SGD is the network hot spot as a single incoming link to a server is shared across multiple workers~\cite{ps-hotspot}. As the PS model scales, using multiple servers alleviates the contention at the server side. However, the number of workers still poses a challenge. In Fire-Caffe~\cite{firecaffe}, a hierarchical tree approach is used to aggregate the gradients from the workers. However, these approaches do not utilize all the underlying communication links and are heavily dependent on the network topology of the machine. Instead, MPI solves the problem by deploying state-of-the-art parallel algorithms which can adapt to any underlying topology. Thus, grouping workers into logical MPI cliques as shown in figure~\ref{fig:ps-mpi-model} significantly resolves contention on PS. 

\textbf{Parameter Staleness:} As the number of workers increases, asynchronous forms of SGD face the issue of staleness~\cite{suyog-staleness,petuum}, which inhibit a fast rate of convergence. Grouping workers into MPI clients potentially offers two immediate advantages: a) It reduces the variance of the gradient updates by effectively increasing the mini\_batch\_size~\cite{leon-variance}. In~\cite{google-sgd}, the number of iterations to converge is halved as the mini\_batch\_size is doubled. b) reduces total number of workers. Depending on the algorithm and the distribution of data, one or both the factors improve the rate. For example, the MPI elastic averaging algorithms studied in this paper benefit from both. Such models offer potential to scale to a full scale machine comprising of thousands of GPUs. 

\textbf{Memory Pressure and Batch Size:} One of the main issues in the implementation of DL systems is memory pressure~\cite{DNN-memory,gpu-ps}, which keeps growing as the number of levels of the network increase. This restricts the choice of batch size of a DL worker to smaller values as the total memory per worker is dictated by the hardware. There are inefficiencies using smaller batch sizes. Grouping workers to larger batches should improve performance as long as the batch sizes falls within algorithmic stipulated limits. Moreover, the new framework also allows the flexibility to decouple dependency between the model mini batch size and memory limits per worker, allowing for a possibility of porting models across different hardware architectures.

We come up with a general framework using MPI and PS that addresses these challenges. The existing approaches have either dealt with one or another but to the best of our knowledge, there is no architecture that addresses all of these in a holistic fashion. 
We use MXNET to demonstrate the new model but the design proposed is flexible and can be adapted to different architectures. MXNET\cite{mxnet} framework, deployed by Amazon in the cloud, is actively developed providing convenient abstractions at many higher level languages \cite{chen2015mxnet,li2014communication}.

\section{DL using MXNET}\label{sec:deep-learning-using}
MXNET uses a declarative computation graph to express the different computations of the DL network and an imperative KVStore to allow for parallelization of deep learning models. 

\subsection{Computation Graph and Execution Model} A declarative model expresses the neural network computation as a graph with all the data flow dependencies. The computation is not immediately executed but rather optimized for efficiency in memory usage and runtime.
Internally, a dependency engine tracks all the operations that can be executed in parallel using the data flow graph. The MXNET dependency engine is generic and can schedule any operation, provided it is tagged with explicit read and write dependencies, shown below. An operation like $a=b+1$ can be performed  by constructing the following lamda function and pushing to the engine:
\texttt{Engine.push(lambda:a.data=b.data+1, read=[b.tag], mutate=[a.tag])}.\\
As we describe in the next section, these lamda functions become key constructs to offload MPI communications and integrate into the data flow graph.


\subsection{KVStore based Aggregation}\label{sec:kvstore-api}
The network parameter and the gradients are expressed as multi-dimensional tensors, implemented as $ndarrays$ in MXNET. Being a data parallel model, the data for a worker is further split across all local GPUs. Hence, there are N such tensors per worker, one per GPU and N GPUs per worker. 
As described in \ref{sec:sgd}, parallel SGD needs to aggregate gradients across all the GPUs before taking the next step. A similar global aggregation needs to be done across all the workers for solving a synchronous SGD. 

\begin{figure}[h]
  \begin{lstlisting}
keys=[1,3,9]
kvstore.init(keys,
             [mxnet.ndarray.ones(shape)]*len(keys))
tensorlist=[mxnet.ndarray.ones(shape,gpu)
            for gpu in gpus]*len(keys)
kvstore.push(keys,tensorlist)
kvstore.pull(keys,tensorlist)
kvstore.set_optimizer(algorithm)
  \end{lstlisting}
  \caption{}
  \label{fig:code-snippet-1}
\end{figure}%
\vspace{-1mm}


The KVStore API provides access to a distributed \emph{\textless{}key,
value\textgreater{}} at the PS. For every mini-batch, the worker computes relevant local updates about the model and uses the KVStore Push and Pull API to synchronize them to a globally visible set of model  variables at PS.
Both primitives take a list of keys and list of values. The python snippet in figure \ref{fig:code-snippet-1} illustrates the semantics
of the operation. As shown in the figure, three keys are initialized in the
KVStore to tensors of a default shape and value (line 2). In the DL algorithm, these keys would correspond to the model weights and initialized to random values using a given distribution. A tensor list is  a succinct representation of the tensors across each level, identified by the key and also one from each GPU of the worker(line4).
The push operation (line 6) first locally aggregates all the tensors of the same key before pushing to the distributed KVStore.
The pull fetches the tensor from the KVStore server corresponding to the given key and copies it to all relevant tensors in tensorlist (line 7). \\
The KVStore API also permits remotely configuring the server to use  different SGD optimizations like momentum SGD, AdaGrad(line 8). We use this to design the elastic averaging technique in MXNET-MPI, described in section ~\ref{sec:kvstore-mpi-algo}.

\begin{figure*}[h]
  \includegraphics[width=\textwidth]{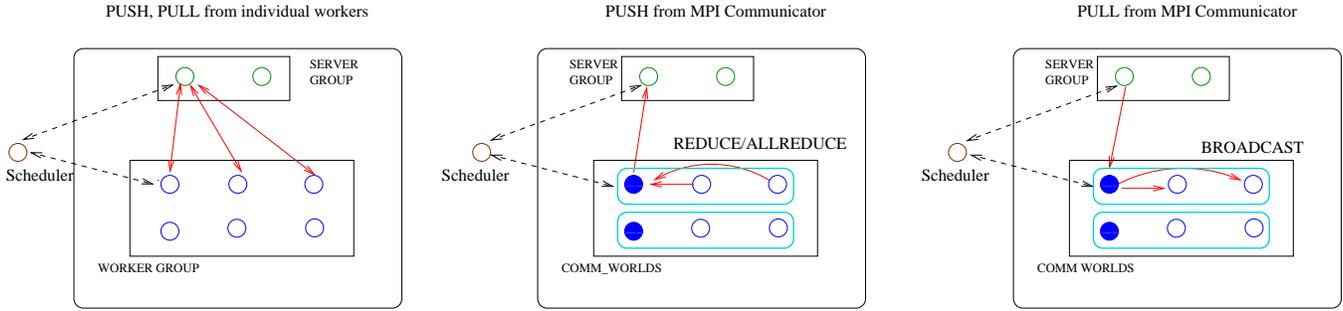}
  \caption{Primitives: push and pull}
  \label{fig:push-pull}
\end{figure*}%



\section{KVStore-MPI Framework}\label{sec:problem}
The basic idea to make the MPI and PS model co-exist is to design a hybrid KVStore-MPI framework extending the existing set of KVStore APIs discussed earlier. In this section we discuss the various design issues to realize the model shown in Figure~\ref{fig:ps-mpi-model} using this new framework.

\subsection{MPI+PS System Architecture}

\subsubsection{Namespaces}
Each worker in the integrated model has identities in two independent namespaces: PS and MPI. In order to do aggregation operations across the workers belonging to the same client, the worker invokes the MPI call usings its MPI rank. For PS updates, it uses its unique name and rank in the PS namespace. In MXNET, each task is either a scheduler, server or a worker as shown in the figure. In the PS namespace,  the scheduler is connected to all the workers and the servers. And each worker is connected to all the servers. MPI Communicators on the other hand are created only among the workers belonging to the same client. We explain how a client is constructed below. 

\subsubsection{Job Launch} We use IBM's LSF~\cite{lsf} administered cluster as our underlying core infrastructure. A launcher is executed on the front end node of the cluster and specified the following parameters: \#workers, \#servers and \#clients. The launcher computes the number of workers in each client using this information. It always launches a MXNET scheduler task first as it listens for all the incoming connections from workers and servers. The address of the scheduler and port information is broad casted to all tasks via the launcher when it spawns them so that connections between the workers and servers can be established.
The launcher uses LSF's bsub command to launch each client as a separate job. In our case, each of these jobs start using mpirun.
LSF transparently manages what hosts to pick the job. Further, \#servers can be tunable, with \#servers equal to zero for pure MPI jobs.


\subsection{Embedding MPI in the Computation Graph}
Except for creation of KVStore, all the other API calls use C++11 lambda functions to "offload" MPI communication to the dependency engine thread in a manner similar to as shown in section ~\ref{sec:deep-learning-using}. The operations are enqueued in order to avoid deadlocks.
\subsubsection{KVStore.create("type")} MPI version of KVStore is constructed by passing appropriate type to the function. Apart from the existing synchronous and asynchronous version supported by MXNET, the additional types would be Synchronous-MPI and Asynchronous-MPI. We provide more details in the following section on how these are used. For MPI, all the workers call MPI\_Init() to create their own MPI\_COMM\_WORLD and become a single MPI client. When there are servers, the rank 0 in PS name space, initializes the values of all the keys on the servers which are subsequently pulled by the rest of the workers. When there are no servers, we use MPI\_Bcast to initialize the weights across all the workers.

\subsubsection{KVStore.push(key, src\_tensor\_list)}In this call, all the workers with rank 0 act as the masters that communicate with the server. In the Push API, tensor allreduce is first used
to aggregate the gradients across the workers in its communicator. The
master then uses the native MXNET C++11 call, ZPush to push the
result to the server. The entire operation is coded as a lambda function and pushed to the engine, along with
the necessary read dependencies shown in Figure~\ref{fig:code-push}. In ~\cite{s-caffe}, the authors also use threaded progress to overlap computation and communication. Their approach is specific to the implementation and doesn't use dependency tracking.

\begin{figure}[h]
  \begin{lstlisting}
auto push_to_servers = [this, key, vals]() {
allreduce(vals.data, vals.len, vals.type, comm);
if(mpi_rank == 0)ZPush(key, vals.data, vals.len, NULL);
}
Engine.Push(push_to_servers, read_deps(vals.tag), mutate(none));
  \end{lstlisting}
  \caption{}
  \label{fig:code-push}
\end{figure}%

\subsubsection{KVStore.pull(key, dst\_tensor\_list)}In the Pull API, the master worker calls broadcast after pulling the value
from the KVStore, as shown in figure~\ref{fig:code-pull}. This is coded as another nested C++11 lambda function passed into the ZPull API for master whose mpi\_rank is zero. Meanwhile, all the others call broadcast.
\begin{figure}[h]
  \begin{lstlisting}
auto pull_from_servers = [this, key, vals]() {
if(mpi_rank == 0)
ZPull(key, vals.data, vals.len,
[vals, comm](){bcast(vals.data, vals.len, vals.type,0, comm)});
else bcast(vals.data, vals.len, vals.type, 0, comm));
};
Engine.Push(pull_from_servers, read_deps(none), mutate(vals.tag));
      \end{lstlisting}
  \caption{}
  \label{fig:code-pull}
\end{figure}%
\subsubsection{KVStore.pushpull(key, src\_tensor\_list, dst\_tensor\_list)}This is a new API added into MXNET with help from MXNET team, that fuses the Push and Pull into one call. This offers a convenient interface amenable to MPI acceleration. Instead of calling the ZPush, ZPull APIs we directly call tensor allreduce, described in section~\ref{sec:tensor-allreduce}.

\section{KVStore-MPI distributed algorithms}\label{sec:kvstore-mpi-algo}
In this section, we show python pseudo codes that each worker in a mpi client executes in the new framework for the three main algorithms discussed in section~\ref{sec:sgd}.\\

\textbf{Synchronous SGD(SGD):}
In MXNET, each worker is assigned a set of data batches, each with a fixed $batch\_size$ which it iterates on. 
The $batch\_size$ is a scheduling unit of MXNET and is different from the $mini\_batch\_size$ of the algorithm. 
As shown in figure~\ref{fig:code-snippet-2}, the executor offloads the declarative computation graph to the engines on line 4. The Push and Pull integrate into this engine by explicitly mentioning the dependencies for scheduling as shown in the previous section.
In the default PS, the servers aggregate all the gradients issued by the worker. However, using MPI the gradients are first aggregated in the respective clients before being pushed to the servers. The $mini\_batch\_size$ for synchronous SGD, is $num\_workers * batch\_size$ (line 10).
\begin{figure}[h]
  \begin{lstlisting}
Kvstore.Create("Synchronous-MPI")
for epoch in range(num_epochs):
 for batch in train_data:
    Executor. Forward_backward(net.symbol, net.params,
                             net.grads, batch) 
    for key in range(num_tensors):
      Kvstore.Push(key, net.grads[key])
      Kvstore.Pull(key, net.grads[key])
    SGD.Update(net.params, net.grads,
                   rescale=1/mini_batch_size)
Executor.ValidationAccuracy(test_batch)
  \end{lstlisting}
  \caption{}
  \label{fig:code-snippet-2}
\end{figure}%
For the pure MPI mode, without servers, the Push and Pull calls are replaced by the
new PushPull API and optimized using tensor allreduce. 

\textbf{Asynchronous SGD(ASGD):}
In this method, the MPI Client executes the asynchronous SGD algorithm and
updates the parameter values as shown in Figure \ref{fig:code-snippet-3}. The specific optimization function is shipped to the server on line 2 along with the scaling associated with the $mini\_batch\_size$. In the MPI version, this is equal to $num\_workers\_per\_client*batch\_size$. The client pushes the locally computed gradients within the communicator and pulls back the new parameter values after the server has finished updating them. 

\begin{figure}[h]
  \begin{lstlisting}
Kvstore.Create("Asynchronous-MPI")
Kvstore.set_optimizer(SGD, rescale=1/mini_batch_size)
for epoch in range(num_epochs):
  for batch in train_data:
    Executor.Forward_backward(net.symbol, net.params,
                                  net.grads, batch)
    for key in range(num_tensors):
      Kvstore.Push(key, net.grads[key])
      Kvstore.Pull(key, net.params[key])
Executor.ValidationAccuracy(test_batch)
  \end{lstlisting}
  \caption{}
  \label{fig:code-snippet-3}
\end{figure}%

\begin{figure}[h]
  \begin{lstlisting}
Kvstore.Create("Asynchronous-MPI")
Kvstore.set_optimizer(Elastic1, rescale=alpha)
iter = 0
for epoch in range(num_epochs):
  for batch in train_data:
    Executor.Forward_backward(net.symbol, net.params,
                                net.grads, batch)
    for key in range(num_tensors):
     if (iter%INTERVAL == 0):
       Kvstore.Push(key, net.params[key])
       Kvstore.Pull(key, net.centers[key])
       Elastic2.(net.params[key], net.centers[key],rescale=alpha)
    SGD.Update(net.params, net.grads, rescale=1/mini_batch_size))
    iter = iter + 1
Executor.ValidationAccuracy(test_batch)
  \end{lstlisting}
  \caption{}
  \label{fig:code-snippet-4}
\end{figure}%
\textbf{Asynchronous Elastic SGD(ESGD):}
As shown in Figure \ref{fig:code-snippet-4}, the MPI client computes SGD within its communicator. The model parameters are pushed/pulled from the server after a certain number of iterations using elastic averaging discussed in section~\ref{sec:parallel-sgd}. Elastic1 is done by the server based on equation ~\ref{elastic1} and it updates the center variables. The second equation~\ref{elastic2} is done by the MPI client using Elastic2 updating its local model parameters. It is shown in
the steps 2 and 12 respectively in the figure. In our experiments, the INTERVAL is set to 64. The $mini\_batch\_size$ is equal to $num\_workers\_per\_client*batch\_size$.
\section{TENSOR COLLECTIVES}\label{sec:tensor-allreduce}
In this section, we describe the design of tensor allreduce that is exclusively used to aggregate gradients from the GPUs across the cluster. 
\subsection{Tensor Operations}We define a tensor to be a grouping of vectors, one from each GPU such that communication operations within a tensor are extremely fast and are expected to scale well as more vectors are added. This is possible because of the growing bandwidths of NVLINK interconnecting the GPUs and launching cuda kernels directly on the GPUs for doing the operation.
For example, the widely used NCCL exclusively uses special kernels to do all the different kinds of operations within the tensor, important being reduce, allreduce, bcast and reduce-scatter. Moreover, for multi-node collectives, thinking about group of vectors as a single unit allows applying several well known collective algorithms that have been used only on one vector at a time. 

This is especially true on the IBM Minsky machines. Figure~\ref{fig:tensor-allreduce-ring} shows the architecture of a Minsky node. Each node consists of two sockets and is directly connected to two NVIDIA P100 Pascal GPUs with independent NVLINK links. The two GPUs are also directly connected to each other using another NVLINK. Essentially, the CPU and the two GPUs in a socket form a 3-clique. Future machines would also allow for a 4-clique configuration where all the compute elements are directly connected to each other using NVLINK. We exploit this property to simultaneously operate on vectors from each GPU, collectively a tensor. 
\subsection{Bucket Algorithms}
 One of the popular choice for designing bandwidth optimized large message collectives is to use bucket algorithms.
 These algorithms use a logical ring which is well known to work in a wide variety of cluster topologies ~\cite{Jain-bucket,allreduce-yuan}. In these algorithms, an allreduce is implemented by a reduce-scatter followed by allgather. It has been proven that these algorithms reach the lower bound for bandwidths. The total cost of the operation is  $ \left(p-1\right))\dot{\alpha}+2\frac{p-1}{p}\dot{n\beta}+\frac{p-1}{p}\dot{n\gamma}$. $\alpha$ is the latency term, $\beta$ is the bandwidth cost per byte and $\gamma$ is the compute cost per byte. $n$ is the total number of bytes used in the operation. For reduce-scatter, the buffer from each process is partitioned into nearly equal parts and after the operation, each process holds a piece of the final reduced sum in its partition.
 
\subsection{Bucket Algorithms on IBM Minsky GPU tensors}
On Minsky nodes, the ring connecting every GPU is not optimal. As shown in the figure, data has to be explicitly copied into the GPU from the host memory as network cannot reach the GPU memory via NVLINK. This would add two extra hops and double the time per ring step. Instead, all our rings use only the host memories. As we show in section~\ref{sec:evaluation}, we obtain very high bandwidths for the tensor reduction and broadcast operations from host memories. Further, the number of hops of the ring is halved as we group the two GPUs from each socket to form a tensor under one worker.

\subsubsection{Tensor Allgather} To design this operation, we re-write the ring algorithm used in OpenMPI~\cite{openmpi} to handle CPU and GPU memories. A pair of buffers is used, one for receiving data from the left neighbor and the other to send data to the right neighbor. In addition to sending the data to its neighbor, we also invoke the broadcast operation from the host buffer to the tensor to overlap the two. Our implementation is organized as a generic GPU tensor library of routines on top of MPI. The tensor library can either use NCCL~\cite{nccl} or custom implementations tailored to IBM's node architecture. For example, to broadcast the tensor on Minsky from host, we invoke two simultaneous cudaMemcpyAsync()~\cite{cuda} calls as the topology allows two different links to the GPUs. 

\subsubsection{Tensor Reduce-Scatter}
\begin{figure*}[h]
\centering
\includegraphics[width=\textwidth]{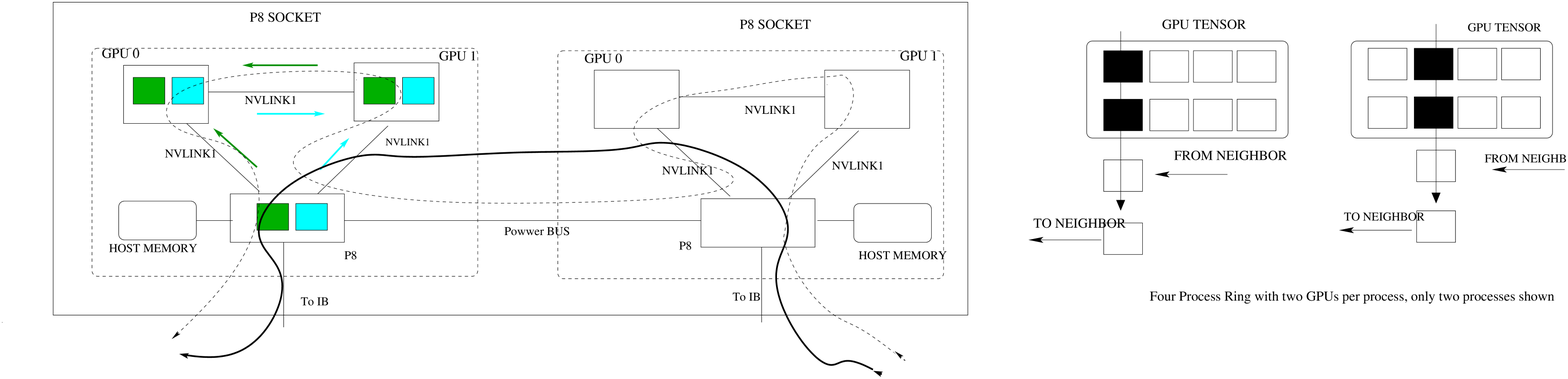}
  \caption{Tensor Allreduce: ring}
  \label{fig:tensor-allreduce-ring}
\end{figure*}

This operation uses the standard bucket algorithm used for buffers in the host memory with one main difference. Instead of reducing a partition of the buffer with the incoming data, a partition of the tensor is used as shown in the figure~\ref{fig:tensor-allreduce-ring}. However, the compute cost now becomes $\frac{\frac{p}{2}-1}{\frac{p}{2}}\dot{n\gamma_{\text{NV}}}$ where $\gamma_{\text{NV}}$ is the reduction cost over NVLINK. We further optimize the operation by overlapping this compute cost with the network transfer. Overlapping this operation within a single ring is not possible as the next communication step of the ring depends on the result of the previous reduction. Thus, we design a multi-ring algorithm where the reduction of the next ring is overlapped with the network transfer of the current ring. The complete tensor\_allreduce operation is described in figure~\ref{fig:allreduce}. The buffer is split equally among the rings and allreduce[ring] operates on the portion of the buffer assigned to it. It uses non blocking GpuStart() routines to launch the CUDA kernels simultaneously with network transfers.
\begin{figure}[h]
  \begin{lstlisting}
uint64_t ring = 0;
allreduce[ring].GpuStart(buffer[ring], step[ring]);
  while(step[NUM_RINGS-2] <= size){
    nextbuf = buffer[(ring+1)%NUM_RINGS];
    nextstep = step[(ring+1)%NUM_RINGS];
    nextring = (ring+1)%NUM_RINGS;
    allreduce[nextring].GpuStart(nextbuf,nextstep);
    buffer[ring] = allreduce[ring].GpuWait();
    if (step[ring] != size){
      buffer[ring] =
      allreduce[ring].SendRecv(buffer[ring],step[ring]);
    }
    step[ring]++;
    ring=(ring+1)%NUM_RINGS;
  }
buffer[ring] = allreduce[ring].GpuWait();
for (unsigned ring = 0; ring < NUM_RINGS; ring++) {
   allreduce[ring].allgather(buffer[ring]);
}
  \end{lstlisting}
  \caption{}
  \label{fig:allreduce}
\end{figure}%
\vspace{-2mm}  



\section{EVALUATION}\label{sec:evaluation}

In this section, we explain the training efficiencies obtained using the new model of MPI + PS on ImageNet 1K. ImageNet 1K contains about 1.2M images with 1000 classes. The total size of the training data used is 336GB and test data is 13GB. We use the latest network, Residual networks(resnet)~\cite{ResNet} using 50 layers for image classification. We ran our tests on two experimental testbeds. Each worker is a process running on a socket and connected to two GPUs.

\textbf{testbed1}: We use a total of 8 dual-socket power8 nodes with 2 Kepler GPUs attached to each socket.The nodes are connected using InfiniBand ConnectX-4 adapters.  Using this testbed, we demonstrate the utility of combining MPI and PS models.
The batch size used is 128 per worker, capped by GPU memory constraints. The following modes of parallelization of model are compared:\\
1)dist-SGD: Uses default PS tasks, all executing synchronous SGD\\
2)dist-ASGD: Uses default PS tasks, all executing asynchronous SGD\\
3)dist-ESGD: Uses default PS tasks, all executing asynchronous ESGD\\
4)mpi-SGD: Uses MPI+PS where gradients are synchronously aggregated first at the MPI clients and next at the PS. \\
5)mpi-ASGD: Uses MPI+PS with gradients synchronously aggregated at the MPI Clients but pushed asynchronously to the PS\\
6)mpi-ESGD: Uses MPI+PS with asynchronous elastic averaging, where the model is computed at the MPI Clients but elastically averaged asynchronously at the PS.\\

\textbf{testbed2}: Comprises of 32 IBM Minsky Power8 nodes with 4 NVIDIA Pascal GPUs on each node connected with InfiniBand CX5 adapters. On this testbed, we demonstrate the tensor collectives and also show the scaling behavior of ImageNet training using the optimizations proposed.



The following metrics are used to measure the performance of our
approaches:

\begin{enumerate}
\def\labelenumi{\alph{enumi})}
\item
  Epoch Time: It's time taken by the workers to train the model over the
  mini-batches of the epoch assigned to it. For multiple workers, we
  take the average time over all the workers.
\item
  Validation Accuracy: The accuracy obtained by using the model on the the separate test samples, done after every epoch. 
\end{enumerate}

\subsection{KVStore-MPI SGD \& ASGD}\label{mpi-sgd-asgdK}
\begin{figure}[h]
\centering
\includegraphics[width=\columnwidth]{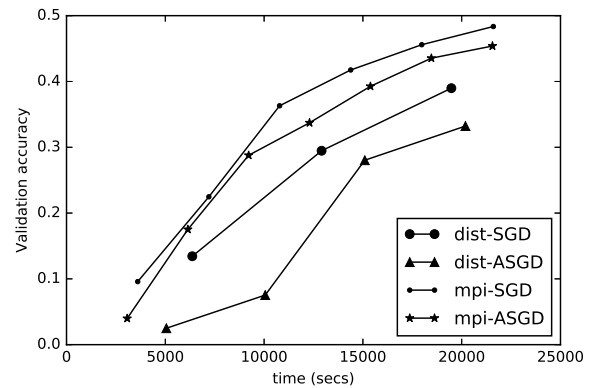}
  \caption{dist-vs-MPI SGD optimizations}
\label{fig:sync-accuracy-epoch}
\end{figure}

\begin{figure}[h]
\centering
\includegraphics[width=\columnwidth]{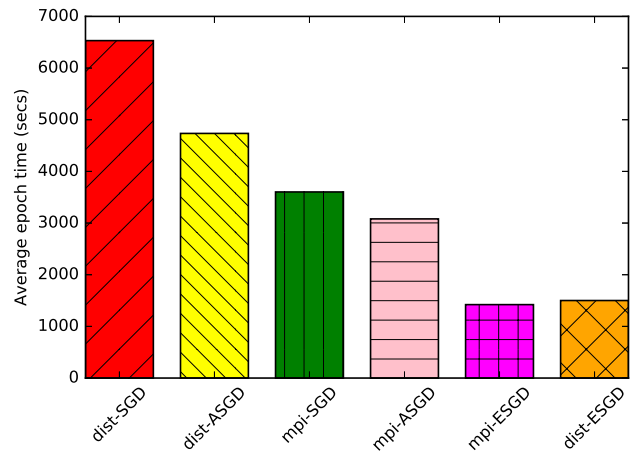}
  \caption{Imagenet Avg Epoch time (seconds)}
  \label{fig:avg-epoch-time-resnet}
\end{figure}

Figure~\ref{fig:sync-accuracy-epoch} shows the ImageNet validation results vs time. On testbed1, we run 12 DL workers two per node using 6 nodes. The two servers are run on the two remaining nodes. From the figure, mpi-SGD trains significantly faster than dist-SGD and mpi-ASGD faster than dist-ASGD. As observed from the figure~\ref{fig:avg-epoch-time-resnet}, using mpi removes the contention, which reduces the time taken for an epoch by the mpi clients vis-a-vis their counterparts. We use two mpi clients with 6 DL workers each interacting with the same two servers. The hot spot is transferred to the mpi client which is better equipped to solve the problem. Also, though the mpi-ASGD is the fastest in the epoch time, it converges slower than mpi-SGD. This is attributed to staleness~\cite{suyog-staleness} where the worker uses parameters from older time steps to compute the gradients. Another factor attributing to the fast convergence of mpi-SGD is the increase in mini\_batch\_size as effectively gradients from 12 workers are averaged which reduces the variance in updates~\cite{leon-variance,google-sgd}. 
Thus, MPI can be used effectively with PS models and the design used in the paper can be replicated in data centers considerably lowering the barrier of adoption of MPI into cloud.

\begin{figure}[h]
\centering
\includegraphics[width=\columnwidth]{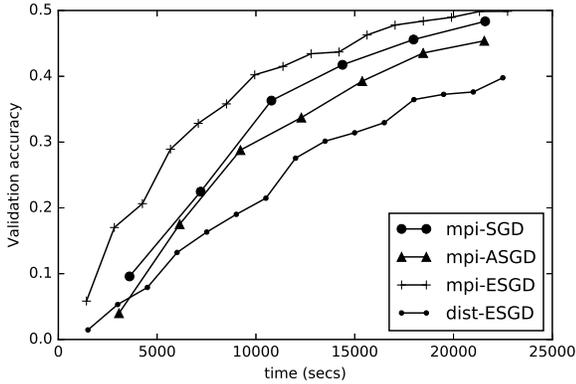}
  \caption{KVStore-MPI based SGD optimizations}
  \label{fig:mpi-sync-vs_mpi-async_vs_mpi-esgd_vs_dist-esgd}
\end{figure}%

\subsection{KVStore-MPI ESGD}\label{sec:mpi-esgd}
\begin{figure}[h]
\centering
\includegraphics[width=\columnwidth]{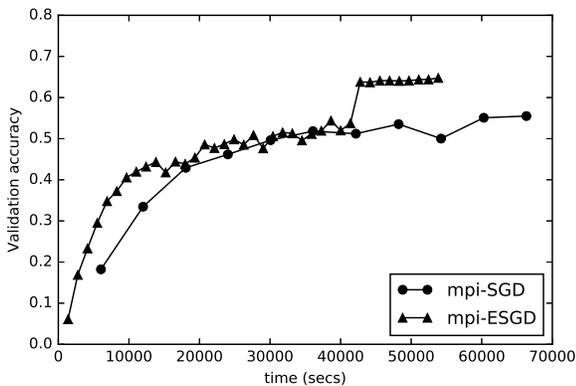}
  \caption{Impact of MPI ESGD}
  \label{fig:mpi-sync_vs_mpi-esgd-time}
\end{figure}

From the previous section, we saw that mpi-SGD is outperforming the rest. However, the scaling of mpi-SGD is dependent upon the communication
bandwidth available for each worker. Even if the problem is weak scaled across nodes keeping batch size the same, the communication cost is expected to increase as the total communication bytes remains the same. After a while, adding extra nodes would offer no benefit~\cite{google-sgd}, as we also see in section~\ref{sec:evaluation}.
The problem is further exacerbated due to the rapid increase in flop count of modern day GPUs~\cite{gpu-computing}.
Thus, having communication avoiding algorithms is extremely helpful if we were to scale cluster wide. ESGD meets this objective. Figure~\ref{fig:mpi-sync-vs_mpi-async_vs_mpi-esgd_vs_dist-esgd} shows the comparison of two ESGD approaches compared to mpi-SGD, mpi-ASGD. The mpi-ESGD approach performs the
best compared to the rest. Here, we use two MPI clients doing independent SGD within the communicators but loosely synchronizing with the PS. Using mpi-ESGD family of protocols provides a path to scale to a full machine. This is seen clearly from figure~\ref{fig:mpi-sync_vs_mpi-esgd-time} with mpi-ESGD out performing mpi-SGD, reaching 0.67 validation accuracy for a multiple epoch run.

It can also be seen that dist-ESGD, with 12 workers is doing the worst of all, in spite of having the same average epoch time as mpi-ESGD with two clients.  
This is because using the mpi-ESGD model allows us to restrict the number of workers and yet allow the scaling of each worker so that staleness is minimized and the optimization algorithm can take advantage of larger mini batch sizes. However, not all application domains and data sets favor increasing the batch size~\cite{largebatch-degrades}, where dist-ESGD would help. This motivates the need for a generic framework as done in this study.

\begin{figure}[h]
\centering
\includegraphics[width=\columnwidth]{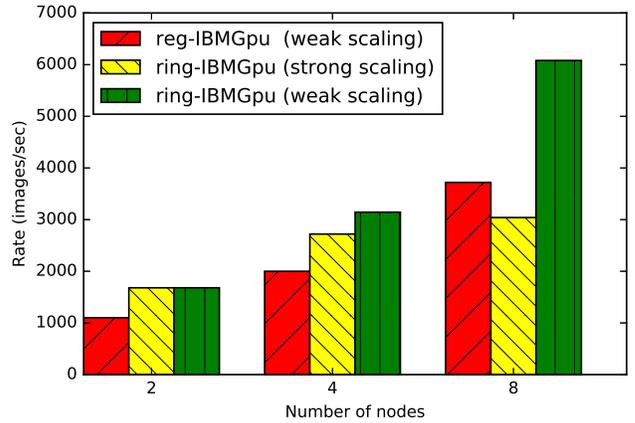}
  \caption{Resnet-50 Scaling behavior}
  \label{fig:resnet-50-scaling}
\end{figure}%
\begin{figure}[h]
\centering
\includegraphics[width=\columnwidth]{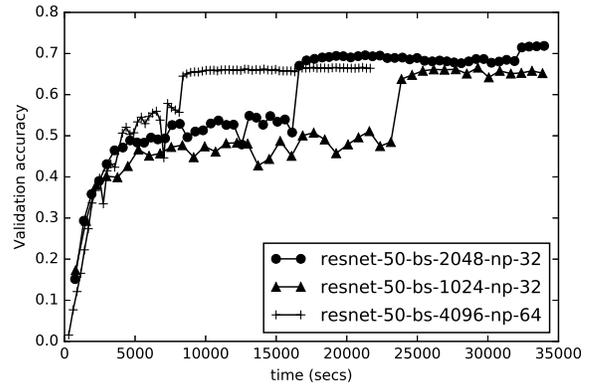}
  \caption{Resnet-50 Learning curves}
  \label{fig:resnet-50-learning-curves}
\end{figure}%

\subsection{Tensor Allreduce}\label{sec:tensor-allreduce}
For doing reductions into host memory, we use either NCCL or IBMGpu, the direct implementation using cuda kernels for doing math. In this approach, the two vectors of the tensor, one on each GPU are split into half and two kernels are launched on the two GPUs to add each in parallel as shown in the figure~\ref{fig:tensor-allreduce-ring}. This allows us to achieve a reduction bandwidth of 30 GB/sec with the final result put back in the host memory. The upper bound would be the write b/w of memory which is 38.4 GB/sec/socket.  With NCCL, we achieve 12 GB/sec with one set of communicators and 15 GB/sec with two sets of communicators. NCCL uses only one thread block where as IBMGpu uses all 112 thread blocks with 1024 threads to keep multiple read/write requests in flight. For broadcast, both IBMGpu and NCCL achieve a b/w of 28 GB/sec. We use the GPU reduction and broadcast routines as a building block to evaluate the different design options of doing tensor allreduce and pick the best among them. The different approaches studied are:
a) ring-IBMGpu design discussed in section~\ref{sec:tensor-allreduce} using two rings,b)ring-NCCL using NCCL with one ring as NCCL operations are blocking in nature, c) omp\_ring-IBMGpu, where the design is similar to the first two, except that the entire buffer is reduced into the host memory and then the host based bucket algorithm is applied with the final results copied back into the GPU. We use 8 OMP threads for data reductions and d) reg-IBMGpu, where the data is reduced into host memory and then the default MPI\_Allreduce is used, followed by a broadcast, with pipelining across the three stages.

\begin{figure}[h]
\centering
\includegraphics[width=\columnwidth]{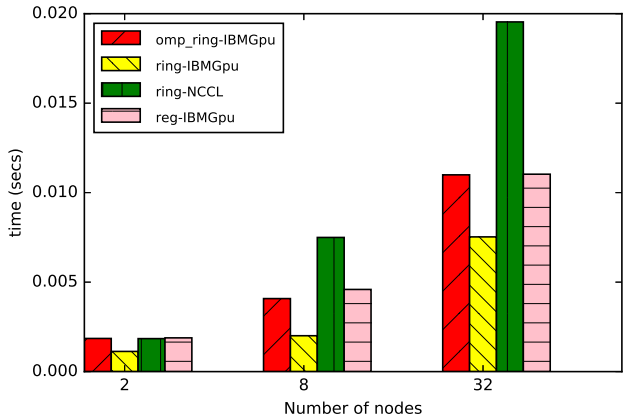}
  \caption{4MB(Message Size)}
  \label{fig:4MB}
\end{figure}

\begin{figure}[h]
\centering
\includegraphics[width=\columnwidth]{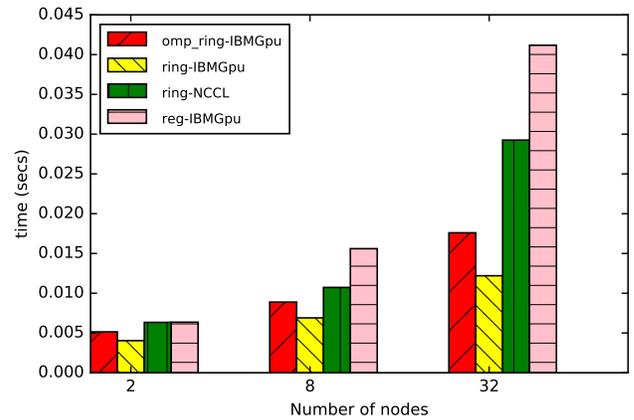}
  \caption{16MB(Message Size)}
  \label{fig:16MB}
\end{figure}

\begin{figure}[h]
\centering
\includegraphics[width=\columnwidth]{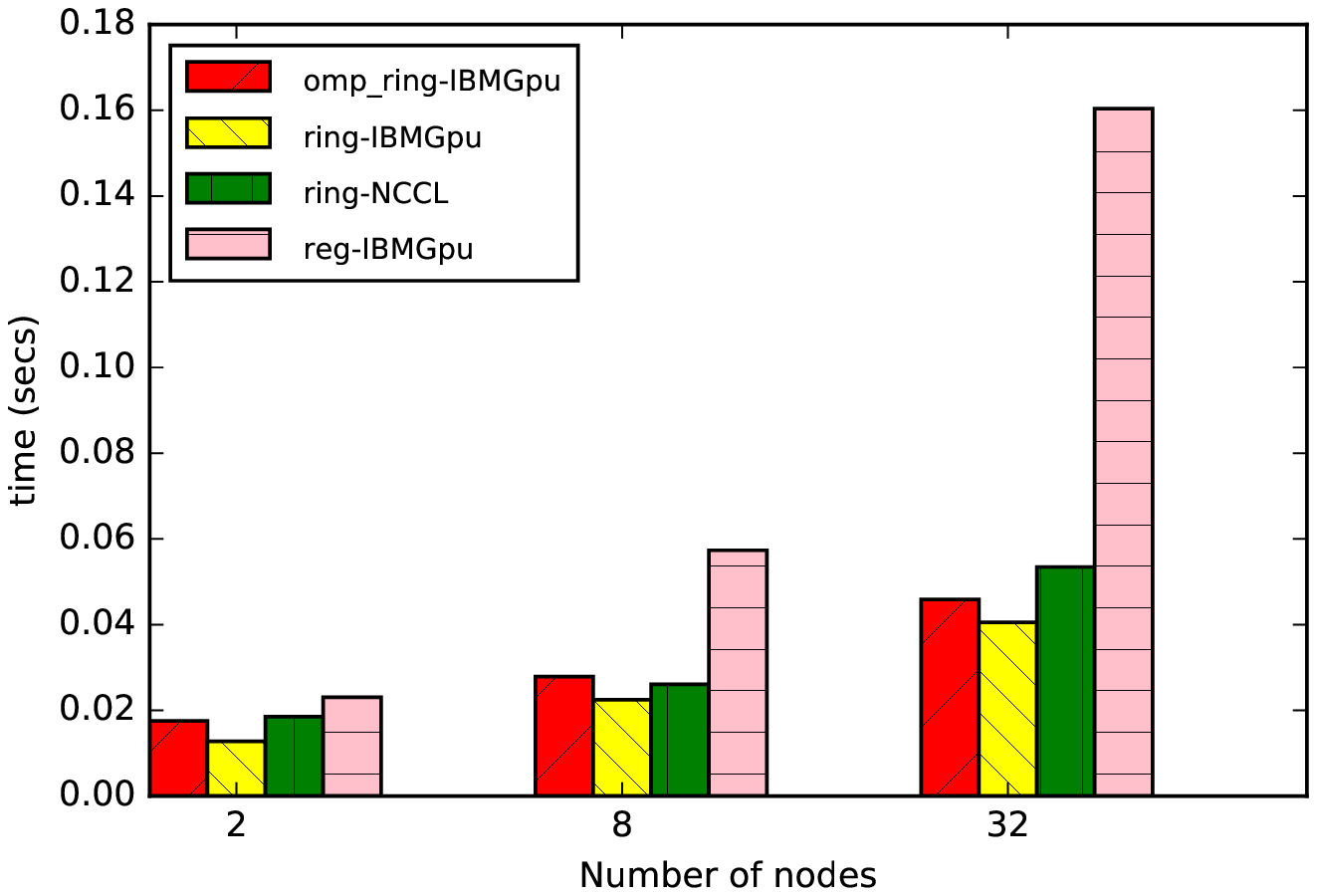}
  \caption{64MB(Message Size)}
  \label{fig:64MB}
\end{figure}

\begin{figure}[h]
\centering
\includegraphics[width=\columnwidth]{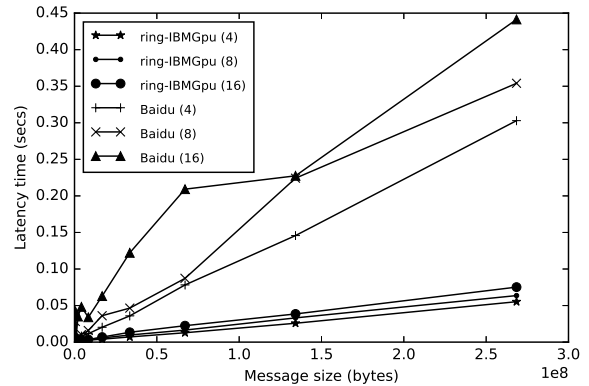}
  \caption{IBMRing-vs-BaiduRing}
  \label{fig:ibm-vs-baidu}
\end{figure}

As shown in figures~\ref{fig:4MB},~\ref{fig:16MB},~\ref{fig:64MB} the IBMGpu ring is doing the best. Apart from using one thread block, another reason for lower NCCL bandwidth is the use of only one NVLINK while doing reduce. For very large messages, the performance gap diminishes across the three as the memory bandwidth becomes the bottleneck. We also see that on Minsky machines we out perform Baidu's ring implementation connecting every GPU, figure~\ref{fig:ibm-vs-baidu} by a factor of six, for the same number of GPUs. We also run large scale ImageNet runs using resnet-50 on testbed2 with \#servers=0, and using mpi-SGD. Our optimizations are nearly twice as fast than using the default, reg-IBMGpu approach, with weak scaling doing the best of all, figure~\ref{fig:resnet-50-scaling}. For strong scaling, the batch size is recursively halved where as it remains constant for weak scaling. 
Figure~\ref{fig:resnet-50-learning-curves} shows the model convergence with our optimized MPI implementation over MXNET and we observe over 0.72 model accuracy, which is the current state-of-the-art. We use an initial learning rate of 0.5 instead of the default 0.1 because of using a larger batch size. 




\section{CONCLUSION}\label{conclusion}
In this paper, we described a hybrid MPI + PS model with a flexibility to allow scaling different DL optimization methods. Though the existing framework supports data parallelism, it can be readily extended to model parallelism. For example, in MPI-ESGD, the parameters can be computed using model parallel optimizations. Also, the framework enables a path for MPI to be useful in cloud based scenarios as it inherits all the task based model attributes needed for fault-tolerance and elasticity. LSF also allows for automatic restart of mpi jobs~\cite{lsfrestart} thus permitting fault recovery. 
Importantly, by embedding MPI into the python modules, the framework allows the user to focus on the algorithms and not deal with explicit MPI parallelization. In this aspect, it is similar to the existing popular frameworks such as Spark~\cite{kim2016deepspark, spark}, Hadoop~\cite{hadoop}.
Moreover, using MPI as the communication glue offers portability and performance for distributed DL optimizations across system and hardware architectures. Finally, the tensor collectives applied to the Minsky architecture are generic and can be readily applied to other GPU based systems.



\bibliographystyle{ACM-Reference-Format}
\bibliography{paper}                     


\begin{thebibliography}{00}


\ifx \showCODEN    \undefined \def \showCODEN     #1{\unskip}     \fi
\ifx \showDOI      \undefined \def \showDOI       #1{{\tt DOI:}\penalty0{#1}\ }
  \fi
\ifx \showISBNx    \undefined \def \showISBNx     #1{\unskip}     \fi
\ifx \showISBNxiii \undefined \def \showISBNxiii  #1{\unskip}     \fi
\ifx \showISSN     \undefined \def \showISSN      #1{\unskip}     \fi
\ifx \showLCCN     \undefined \def \showLCCN      #1{\unskip}     \fi
\ifx \shownote     \undefined \def \shownote      #1{#1}          \fi
\ifx \showarticletitle \undefined \def \showarticletitle #1{#1}   \fi
\ifx \showURL      \undefined \def \showURL       #1{#1}          \fi
\providecommand\bibfield[2]{#2}
\providecommand\bibinfo[2]{#2}
\providecommand\natexlab[1]{#1}
\providecommand\showeprint[2][]{arXiv:#2}

\bibitem[\protect\citeauthoryear{??}{cor}{2016}]%
        {coral}
 \bibinfo{year}{2016}\natexlab{}.
\newblock \bibinfo{title}{CORAL}.
\newblock
  \bibinfo{howpublished}{\url{http://www.nextplatform.com/2016/04/04/eyes-ibm-future-supercomputing-push}}.
    (\bibinfo{year}{2016}).
\newblock


\bibitem[\protect\citeauthoryear{??}{aws}{2017}]%
        {aws}
 \bibinfo{year}{2017}\natexlab{}.
\newblock \bibinfo{title}{AWS}.
\newblock \bibinfo{howpublished}{\url{https://aws.amazon.com/}}.
  (\bibinfo{year}{2017}).
\newblock


\bibitem[\protect\citeauthoryear{??}{azu}{2017}]%
        {azure}
 \bibinfo{year}{2017}\natexlab{}.
\newblock \bibinfo{title}{Azure}.
\newblock
  \bibinfo{howpublished}{\url{http://www.nvidia.com/object/gpu-accelerated-microsoft-azure.html}}.
    (\bibinfo{year}{2017}).
\newblock


\bibitem[\protect\citeauthoryear{??}{CNT}{2017}]%
        {CNTK}
 \bibinfo{year}{2017}\natexlab{}.
\newblock \bibinfo{title}{CNTK}.
\newblock \bibinfo{howpublished}{\url{https://github.com/Microsoft/CNTK}}.
  (\bibinfo{year}{2017}).
\newblock


\bibitem[\protect\citeauthoryear{??}{cud}{2017}]%
        {cuda}
 \bibinfo{year}{2017}\natexlab{}.
\newblock \bibinfo{title}{CUDA}.
\newblock
  \bibinfo{howpublished}{\url{http://horacio9573.no-ip.org/cuda/index.html}}.
  (\bibinfo{year}{2017}).
\newblock


\bibitem[\protect\citeauthoryear{??}{goo}{2017}]%
        {google-cloud}
 \bibinfo{year}{2017}\natexlab{}.
\newblock \bibinfo{title}{Google Cloud}.
\newblock \bibinfo{howpublished}{\url{https://cloud.google.com/gpu/}}.
  (\bibinfo{year}{2017}).
\newblock


\bibitem[\protect\citeauthoryear{??}{had}{2017}]%
        {hadoop}
 \bibinfo{year}{2017}\natexlab{}.
\newblock \bibinfo{title}{Hadoop}.
\newblock \bibinfo{howpublished}{\url{http://hadoop.apache.org/}}.
  (\bibinfo{year}{2017}).
\newblock


\bibitem[\protect\citeauthoryear{??}{ibm}{2017a}]%
        {ibm-cloud}
 \bibinfo{year}{2017}\natexlab{a}.
\newblock \bibinfo{title}{IBM Cloud}.
\newblock \bibinfo{howpublished}{\url{https://www.ibm.com/cloud-computing/}}.
  (\bibinfo{year}{2017}).
\newblock


\bibitem[\protect\citeauthoryear{??}{ibm}{2017b}]%
        {ibm-minsky}
 \bibinfo{year}{2017}\natexlab{b}.
\newblock \bibinfo{title}{IBM Minsky}.
\newblock
  \bibinfo{howpublished}{\url{https://www.hpcwire.com/2016/09/08/ibm-debuts-power8-chip-nvlink-3-new-systems/}}.
    (\bibinfo{year}{2017}).
\newblock


\bibitem[\protect\citeauthoryear{??}{lsf}{2017a}]%
        {lsf}
 \bibinfo{year}{2017}\natexlab{a}.
\newblock \bibinfo{title}{LSF}.
\newblock
  \bibinfo{howpublished}{\url{https://www.ibm.com/support/knowledgecenter/SSETD4_9.1.3/lsf_users_guide/chap_lsf_about.html}}.
    (\bibinfo{year}{2017}).
\newblock


\bibitem[\protect\citeauthoryear{??}{lsf}{2017b}]%
        {lsfrestart}
 \bibinfo{year}{2017}\natexlab{b}.
\newblock \bibinfo{title}{LSF restarting job}.
\newblock
  \bibinfo{howpublished}{\url{https://www.ibm.com/support/knowledgecenter/SSETD4_9.1.3/lsf_admin/job_requeue_auto_config.html}}.
    (\bibinfo{year}{2017}).
\newblock


\bibitem[\protect\citeauthoryear{??}{caf}{2017}]%
        {caffeMPIindiana}
 \bibinfo{year}{2017}\natexlab{}.
\newblock \bibinfo{title}{mpi-caffe}.
\newblock
  \bibinfo{howpublished}{\url{https://computing.ece.vt.edu/~steflee/mpi-caffe.html}}.
    (\bibinfo{year}{2017}).
\newblock


\bibitem[\protect\citeauthoryear{??}{mpi}{2017}]%
        {mpi-dynamic-issues}
 \bibinfo{year}{2017}\natexlab{}.
\newblock \bibinfo{title}{MPI Resize}.
\newblock
  \bibinfo{howpublished}{\url{https://www.arcos.inf.uc3m.es/wp-content/uploads/sites/47/2017/02/wg3-E3.1-v1.0.pdf}}.
    (\bibinfo{year}{2017}).
\newblock


\bibitem[\protect\citeauthoryear{??}{mxn}{2017}]%
        {mxnet}
 \bibinfo{year}{2017}\natexlab{}.
\newblock \bibinfo{title}{MXNET}.
\newblock \bibinfo{howpublished}{\url{http://mxnet.io/}}.
  (\bibinfo{year}{2017}).
\newblock


\bibitem[\protect\citeauthoryear{??}{ncc}{2017}]%
        {nccl}
 \bibinfo{year}{2017}\natexlab{}.
\newblock \bibinfo{title}{NCCL}.
\newblock \bibinfo{howpublished}{\url{https://github.com/NVIDIA/nccl}}.
  (\bibinfo{year}{2017}).
\newblock


\bibitem[\protect\citeauthoryear{??}{ope}{2017}]%
        {openmpi}
 \bibinfo{year}{2017}\natexlab{}.
\newblock \bibinfo{title}{OpenMPI}.
\newblock \bibinfo{howpublished}{\url{https://www.open-mpi.org/}}.
  (\bibinfo{year}{2017}).
\newblock


\bibitem[\protect\citeauthoryear{??}{spa}{2017}]%
        {spark}
 \bibinfo{year}{2017}\natexlab{}.
\newblock \bibinfo{title}{Spark}.
\newblock \bibinfo{howpublished}{\url{http://spark.apache.org/}}.
  (\bibinfo{year}{2017}).
\newblock


\bibitem[\protect\citeauthoryear{??}{ulf}{2017}]%
        {ulfm}
 \bibinfo{year}{2017}\natexlab{}.
\newblock \bibinfo{title}{ULFM}.
\newblock \bibinfo{howpublished}{\url{http://fault-tolerance.org/}}.
  (\bibinfo{year}{2017}).
\newblock


\bibitem[\protect\citeauthoryear{Abadi, Barham, Chen, Chen, Davis, Dean, Devin,
  Ghemawat, Irving, Isard, et~al\mbox{.}}{Abadi et~al\mbox{.}}{2016}]%
        {abadi2016tensorflow}
\bibfield{author}{\bibinfo{person}{Mart{\'\i}n Abadi}, \bibinfo{person}{Paul
  Barham}, \bibinfo{person}{Jianmin Chen}, \bibinfo{person}{Zhifeng Chen},
  \bibinfo{person}{Andy Davis}, \bibinfo{person}{Jeffrey Dean},
  \bibinfo{person}{Matthieu Devin}, \bibinfo{person}{Sanjay Ghemawat},
  \bibinfo{person}{Geoffrey Irving}, \bibinfo{person}{Michael Isard}, {and}
  \bibinfo{person}{others}.} \bibinfo{year}{2016}\natexlab{}.
\newblock \showarticletitle{TensorFlow: A system for large-scale machine
  learning}.
\newblock \bibinfo{journal}{{\em arXiv preprint arXiv:1605.08695\/}}
  (\bibinfo{year}{2016}).
\newblock


\bibitem[\protect\citeauthoryear{Awan, Hamidouche, Hashmi, and Panda}{Awan
  et~al\mbox{.}}{2017}]%
        {s-caffe}
\bibfield{author}{\bibinfo{person}{Ammar~Ahmad Awan}, \bibinfo{person}{Khaled
  Hamidouche}, \bibinfo{person}{Jahanzeb~Maqbool Hashmi}, {and}
  \bibinfo{person}{Dhabaleswar~K. Panda}.} \bibinfo{year}{2017}\natexlab{}.
\newblock \showarticletitle{S-Caffe: Co-designing MPI Runtimes and Caffe for
  Scalable Deep Learning on Modern GPU Clusters}. In \bibinfo{booktitle}{{\em
  Proceedings of the 22Nd ACM SIGPLAN Symposium on Principles and Practice of
  Parallel Programming}} {\em (\bibinfo{series}{PPoPP '17})}.
  \bibinfo{publisher}{ACM}, \bibinfo{address}{New York, NY, USA},
  \bibinfo{pages}{193--205}.
\newblock
\showISBNx{978-1-4503-4493-7}
\showDOI{%
\url{https://doi.org/10.1145/3018743.3018769}}


\bibitem[\protect\citeauthoryear{Bengio}{Bengio}{2012}]%
        {bengio2012practical}
\bibfield{author}{\bibinfo{person}{Yoshua Bengio}.}
  \bibinfo{year}{2012}\natexlab{}.
\newblock \showarticletitle{Practical recommendations for gradient-based
  training of deep architectures}.
\newblock \bibinfo{journal}{{\em CoRR\/}}  \bibinfo{volume}{abs/1206.5533}
  (\bibinfo{year}{2012}).
\newblock
\showURL{%
\url{http://arxiv.org/abs/1206.5533}}


\bibitem[\protect\citeauthoryear{Bland, Bouteiller, H{\'{e}}rault, Hursey,
  Bosilca, and Dongarra}{Bland et~al\mbox{.}}{2013}]%
        {openmpi-ulfm}
\bibfield{author}{\bibinfo{person}{Wesley Bland}, \bibinfo{person}{Aurelien
  Bouteiller}, \bibinfo{person}{Thomas H{\'{e}}rault}, \bibinfo{person}{Joshua
  Hursey}, \bibinfo{person}{George Bosilca}, {and} \bibinfo{person}{Jack~J.
  Dongarra}.} \bibinfo{year}{2013}\natexlab{}.
\newblock \showarticletitle{An evaluation of User-Level Failure Mitigation
  support in {MPI}}.
\newblock \bibinfo{journal}{{\em Computing\/}} \bibinfo{volume}{95},
  \bibinfo{number}{12} (\bibinfo{year}{2013}), \bibinfo{pages}{1171--1184}.
\newblock
\showDOI{%
\url{https://doi.org/10.1007/s00607-013-0331-3}}


\bibitem[\protect\citeauthoryear{Bland, Lu, Seo, and Balaji}{Bland
  et~al\mbox{.}}{2015}]%
        {mpich-ulfm}
\bibfield{author}{\bibinfo{person}{Wesley Bland}, \bibinfo{person}{Huiwei Lu},
  \bibinfo{person}{Sangmin Seo}, {and} \bibinfo{person}{Pavan Balaji}.}
  \bibinfo{year}{2015}\natexlab{}.
\newblock \showarticletitle{Lessons Learned Implementing User-Level Failure
  Mitigation in {MPICH}}. In \bibinfo{booktitle}{{\em 15th {IEEE/ACM}
  International Symposium on Cluster, Cloud and Grid Computing, CCGrid 2015,
  Shenzhen, China, May 4-7, 2015}}. \bibinfo{pages}{1123--1126}.
\newblock
\showDOI{%
\url{https://doi.org/10.1109/CCGrid.2015.51}}


\bibitem[\protect\citeauthoryear{{Bottou}, {Curtis}, and {Nocedal}}{{Bottou}
  et~al\mbox{.}}{2016}]%
        {leon-variance}
\bibfield{author}{\bibinfo{person}{L. {Bottou}}, \bibinfo{person}{F.~E.
  {Curtis}}, {and} \bibinfo{person}{J. {Nocedal}}.}
  \bibinfo{year}{2016}\natexlab{}.
\newblock \showarticletitle{{Optimization Methods for Large-Scale Machine
  Learning}}.
\newblock \bibinfo{journal}{{\em ArXiv e-prints\/}} (\bibinfo{date}{June}
  \bibinfo{year}{2016}).
\newblock
\showeprint[arxiv]{stat.ML/1606.04838}


\bibitem[\protect\citeauthoryear{Cao, Zhang, Joachims, Webb, Margineantu, and
  Williams}{Cao et~al\mbox{.}}{2015}]%
        {petuum}
\bibfield{editor}{\bibinfo{person}{Longbing Cao}, \bibinfo{person}{Chengqi
  Zhang}, \bibinfo{person}{Thorsten Joachims}, \bibinfo{person}{Geoffrey~I.
  Webb}, \bibinfo{person}{Dragos~D. Margineantu}, {and} \bibinfo{person}{Graham
  Williams}} (Eds.). \bibinfo{year}{2015}\natexlab{}.
\newblock \bibinfo{booktitle}{{\em Proceedings of the 21th {ACM} {SIGKDD}
  International Conference on Knowledge Discovery and Data Mining, Sydney, NSW,
  Australia, August 10-13, 2015}}. \bibinfo{publisher}{{ACM}}.
\newblock
\showISBNx{978-1-4503-3664-2}
\showURL{%
\url{http://dl.acm.org/citation.cfm?id=2783258}}


\bibitem[\protect\citeauthoryear{Chen, Li, Li, Lin, Wang, Wang, Xiao, Xu,
  Zhang, and Zhang}{Chen et~al\mbox{.}}{2015}]%
        {chen2015mxnet}
\bibfield{author}{\bibinfo{person}{Tianqi Chen}, \bibinfo{person}{Mu Li},
  \bibinfo{person}{Yutian Li}, \bibinfo{person}{Min Lin},
  \bibinfo{person}{Naiyan Wang}, \bibinfo{person}{Minjie Wang},
  \bibinfo{person}{Tianjun Xiao}, \bibinfo{person}{Bing Xu},
  \bibinfo{person}{Chiyuan Zhang}, {and} \bibinfo{person}{Zheng Zhang}.}
  \bibinfo{year}{2015}\natexlab{}.
\newblock \showarticletitle{Mxnet: A flexible and efficient machine learning
  library for heterogeneous distributed systems}.
\newblock \bibinfo{journal}{{\em arXiv preprint arXiv:1512.01274\/}}
  (\bibinfo{year}{2015}).
\newblock


\bibitem[\protect\citeauthoryear{Chilimbi, Suzue, Apacible, and
  Kalyanaraman}{Chilimbi et~al\mbox{.}}{2014}]%
        {ps-hotspot}
\bibfield{author}{\bibinfo{person}{Trishul Chilimbi}, \bibinfo{person}{Yutaka
  Suzue}, \bibinfo{person}{Johnson Apacible}, {and} \bibinfo{person}{Karthik
  Kalyanaraman}.} \bibinfo{year}{2014}\natexlab{}.
\newblock \showarticletitle{Project Adam: Building an Efficient and Scalable
  Deep Learning Training System}. In \bibinfo{booktitle}{{\em 11th USENIX
  Symposium on Operating Systems Design and Implementation (OSDI 14)}}.
  \bibinfo{publisher}{USENIX Association}, \bibinfo{address}{Broomfield, CO},
  \bibinfo{pages}{571--582}.
\newblock
\showISBNx{978-1-931971-16-4}
\showURL{%
\url{https://www.usenix.org/conference/osdi14/technical-sessions/presentation/chilimbi}}


\bibitem[\protect\citeauthoryear{Coates, Huval, Wang, Wu, Ng, and
  Catanzaro}{Coates et~al\mbox{.}}{2013}]%
        {dl-cots}
\bibfield{author}{\bibinfo{person}{Adam Coates}, \bibinfo{person}{Brody Huval},
  \bibinfo{person}{Tao Wang}, \bibinfo{person}{David~J. Wu},
  \bibinfo{person}{Andrew~Y. Ng}, {and} \bibinfo{person}{Bryan Catanzaro}.}
  \bibinfo{year}{2013}\natexlab{}.
\newblock \showarticletitle{Deep Learning with COTS HPC Systems}. In
  \bibinfo{booktitle}{{\em Proceedings of the 30th International Conference on
  International Conference on Machine Learning - Volume 28}} {\em
  (\bibinfo{series}{ICML'13})}. \bibinfo{publisher}{JMLR.org},
  \bibinfo{pages}{III--1337--III--1345}.
\newblock
\showURL{%
\url{http://dl.acm.org/citation.cfm?id=3042817.3043086}}


\bibitem[\protect\citeauthoryear{Cui, Zhang, Ganger, Gibbons, and Xing}{Cui
  et~al\mbox{.}}{2016}]%
        {gpu-ps}
\bibfield{author}{\bibinfo{person}{Henggang Cui}, \bibinfo{person}{Hao Zhang},
  \bibinfo{person}{Gregory~R. Ganger}, \bibinfo{person}{Phillip~B. Gibbons},
  {and} \bibinfo{person}{Eric~P. Xing}.} \bibinfo{year}{2016}\natexlab{}.
\newblock \showarticletitle{GeePS: Scalable Deep Learning on Distributed GPUs
  with a GPU-specialized Parameter Server}. In \bibinfo{booktitle}{{\em
  Proceedings of the Eleventh European Conference on Computer Systems}} {\em
  (\bibinfo{series}{EuroSys '16})}. \bibinfo{publisher}{ACM},
  \bibinfo{address}{New York, NY, USA}, Article \bibinfo{articleno}{4},
  \bibinfo{numpages}{16}~pages.
\newblock
\showISBNx{978-1-4503-4240-7}
\showDOI{%
\url{https://doi.org/10.1145/2901318.2901323}}


\bibitem[\protect\citeauthoryear{Das, Avancha, Mudigere, Vaidyanathan,
  Sridharan, Kalamkar, Kaul, and Dubey}{Das et~al\mbox{.}}{2016}]%
        {dl-intel}
\bibfield{author}{\bibinfo{person}{Dipankar Das}, \bibinfo{person}{Sasikanth
  Avancha}, \bibinfo{person}{Dheevatsa Mudigere}, \bibinfo{person}{Karthikeyan
  Vaidyanathan}, \bibinfo{person}{Srinivas Sridharan},
  \bibinfo{person}{Dhiraj~D. Kalamkar}, \bibinfo{person}{Bharat Kaul}, {and}
  \bibinfo{person}{Pradeep Dubey}.} \bibinfo{year}{2016}\natexlab{}.
\newblock \showarticletitle{Distributed Deep Learning Using Synchronous
  Stochastic Gradient Descent}.
\newblock \bibinfo{journal}{{\em CoRR\/}}  \bibinfo{volume}{abs/1602.06709}
  (\bibinfo{year}{2016}).
\newblock
\showURL{%
\url{http://arxiv.org/abs/1602.06709}}


\bibitem[\protect\citeauthoryear{Dean, Corrado, Monga, Chen, Devin, Le, Mao,
  Ranzato, Senior, Tucker, Yang, and Ng}{Dean et~al\mbox{.}}{2012}]%
        {google-DNN}
\bibfield{author}{\bibinfo{person}{Jeffrey Dean}, \bibinfo{person}{Greg~S.
  Corrado}, \bibinfo{person}{Rajat Monga}, \bibinfo{person}{Kai Chen},
  \bibinfo{person}{Matthieu Devin}, \bibinfo{person}{Quoc~V. Le},
  \bibinfo{person}{Mark~Z. Mao}, \bibinfo{person}{Marc'Aurelio Ranzato},
  \bibinfo{person}{Andrew Senior}, \bibinfo{person}{Paul Tucker},
  \bibinfo{person}{Ke Yang}, {and} \bibinfo{person}{Andrew~Y. Ng}.}
  \bibinfo{year}{2012}\natexlab{}.
\newblock \showarticletitle{Large Scale Distributed Deep Networks}. In
  \bibinfo{booktitle}{{\em Proceedings of the 25th International Conference on
  Neural Information Processing Systems}} {\em (\bibinfo{series}{NIPS'12})}.
  \bibinfo{publisher}{Curran Associates Inc.}, \bibinfo{address}{USA},
  \bibinfo{pages}{1223--1231}.
\newblock
\showURL{%
\url{http://dl.acm.org/citation.cfm?id=2999134.2999271}}


\bibitem[\protect\citeauthoryear{He, Zhang, Ren, and Sun}{He
  et~al\mbox{.}}{2015}]%
        {ResNet}
\bibfield{author}{\bibinfo{person}{Kaiming He}, \bibinfo{person}{Xiangyu
  Zhang}, \bibinfo{person}{Shaoqing Ren}, {and} \bibinfo{person}{Jian Sun}.}
  \bibinfo{year}{2015}\natexlab{}.
\newblock \showarticletitle{Deep Residual Learning for Image Recognition}.
\newblock \bibinfo{journal}{{\em CoRR\/}}  \bibinfo{volume}{abs/1512.03385}
  (\bibinfo{year}{2015}).
\newblock
\showURL{%
\url{http://arxiv.org/abs/1512.03385}}


\bibitem[\protect\citeauthoryear{Iandola, Ashraf, Moskewicz, and
  Keutzer}{Iandola et~al\mbox{.}}{2015}]%
        {firecaffe}
\bibfield{author}{\bibinfo{person}{Forrest~N. Iandola}, \bibinfo{person}{Khalid
  Ashraf}, \bibinfo{person}{Matthew~W. Moskewicz}, {and} \bibinfo{person}{Kurt
  Keutzer}.} \bibinfo{year}{2015}\natexlab{}.
\newblock \showarticletitle{FireCaffe: near-linear acceleration of deep neural
  network training on compute clusters.}
\newblock \bibinfo{journal}{{\em CoRR\/}}  \bibinfo{volume}{abs/1511.00175}
  (\bibinfo{year}{2015}).
\newblock
\showURL{%
\url{http://dblp.uni-trier.de/db/journals/corr/corr1511.html\#IandolaAMK15}}


\bibitem[\protect\citeauthoryear{Jain and Sabharwal}{Jain and
  Sabharwal}{2010}]%
        {Jain-bucket}
\bibfield{author}{\bibinfo{person}{Nikhil Jain} {and} \bibinfo{person}{Yogish
  Sabharwal}.} \bibinfo{year}{2010}\natexlab{}.
\newblock \showarticletitle{Optimal Bucket Algorithms for Large MPI Collectives
  on Torus Interconnects}. In \bibinfo{booktitle}{{\em Proceedings of the 24th
  ACM International Conference on Supercomputing}} {\em (\bibinfo{series}{ICS
  '10})}. \bibinfo{publisher}{ACM}, \bibinfo{address}{New York, NY, USA},
  \bibinfo{pages}{27--36}.
\newblock
\showISBNx{978-1-4503-0018-6}
\showDOI{%
\url{https://doi.org/10.1145/1810085.1810093}}


\bibitem[\protect\citeauthoryear{Keskar, Mudigere, Nocedal, Smelyanskiy, and
  Tang}{Keskar et~al\mbox{.}}{2016}]%
        {largebatch-degrades}
\bibfield{author}{\bibinfo{person}{Nitish~Shirish Keskar},
  \bibinfo{person}{Dheevatsa Mudigere}, \bibinfo{person}{Jorge Nocedal},
  \bibinfo{person}{Mikhail Smelyanskiy}, {and} \bibinfo{person}{Ping Tak~Peter
  Tang}.} \bibinfo{year}{2016}\natexlab{}.
\newblock \showarticletitle{On Large-Batch Training for Deep Learning:
  Generalization Gap and Sharp Minima}.
\newblock \bibinfo{journal}{{\em CoRR\/}}  \bibinfo{volume}{abs/1609.04836}
  (\bibinfo{year}{2016}).
\newblock
\showURL{%
\url{http://arxiv.org/abs/1609.04836}}


\bibitem[\protect\citeauthoryear{Kim, Park, Jang, and Yoon}{Kim
  et~al\mbox{.}}{2016}]%
        {kim2016deepspark}
\bibfield{author}{\bibinfo{person}{Hanjoo Kim}, \bibinfo{person}{Jaehong Park},
  \bibinfo{person}{Jaehee Jang}, {and} \bibinfo{person}{Sungroh Yoon}.}
  \bibinfo{year}{2016}\natexlab{}.
\newblock \showarticletitle{DeepSpark: Spark-Based Deep Learning Supporting
  Asynchronous Updates and Caffe Compatibility}.
\newblock \bibinfo{journal}{{\em arXiv preprint arXiv:1602.08191\/}}
  (\bibinfo{year}{2016}).
\newblock


\bibitem[\protect\citeauthoryear{Laguna, Richards, Gamblin, Schulz, and
  de~Supinski}{Laguna et~al\mbox{.}}{2014}]%
        {ulfm-issues}
\bibfield{author}{\bibinfo{person}{Ignacio Laguna}, \bibinfo{person}{David~F.
  Richards}, \bibinfo{person}{Todd Gamblin}, \bibinfo{person}{Martin Schulz},
  {and} \bibinfo{person}{Bronis~R. de Supinski}.}
  \bibinfo{year}{2014}\natexlab{}.
\newblock \showarticletitle{Evaluating User-Level Fault Tolerance for MPI
  Applications}. In \bibinfo{booktitle}{{\em Proceedings of the 21st European
  MPI Users' Group Meeting}} {\em (\bibinfo{series}{EuroMPI/ASIA '14})}.
  \bibinfo{publisher}{ACM}, \bibinfo{address}{New York, NY, USA}, Article
  \bibinfo{articleno}{57}, \bibinfo{numpages}{6}~pages.
\newblock
\showISBNx{978-1-4503-2875-3}
\showDOI{%
\url{https://doi.org/10.1145/2642769.2642775}}


\bibitem[\protect\citeauthoryear{Li, Andersen, Smola, and Yu}{Li
  et~al\mbox{.}}{2014}]%
        {li2014communication}
\bibfield{author}{\bibinfo{person}{Mu Li}, \bibinfo{person}{David~G Andersen},
  \bibinfo{person}{Alex~J Smola}, {and} \bibinfo{person}{Kai Yu}.}
  \bibinfo{year}{2014}\natexlab{}.
\newblock \showarticletitle{Communication efficient distributed machine
  learning with the parameter server}. In \bibinfo{booktitle}{{\em Advances in
  Neural Information Processing Systems}}. \bibinfo{pages}{19--27}.
\newblock


\bibitem[\protect\citeauthoryear{Lian, Huang, Li, and Liu}{Lian
  et~al\mbox{.}}{2015}]%
        {lian2015asynchronous}
\bibfield{author}{\bibinfo{person}{Xiangru Lian}, \bibinfo{person}{Yijun
  Huang}, \bibinfo{person}{Yuncheng Li}, {and} \bibinfo{person}{Ji Liu}.}
  \bibinfo{year}{2015}\natexlab{}.
\newblock \showarticletitle{Asynchronous parallel stochastic gradient for
  nonconvex optimization}. In \bibinfo{booktitle}{{\em Advances in Neural
  Information Processing Systems}}. \bibinfo{pages}{2737--2745}.
\newblock


\bibitem[\protect\citeauthoryear{McMahan, Moore, Ramage, and y~Arcas}{McMahan
  et~al\mbox{.}}{2016}]%
        {federated-averaging}
\bibfield{author}{\bibinfo{person}{H.~Brendan McMahan}, \bibinfo{person}{Eider
  Moore}, \bibinfo{person}{Daniel Ramage}, {and} \bibinfo{person}{Blaise~Aguera
  y Arcas}.} \bibinfo{year}{2016}\natexlab{}.
\newblock \bibinfo{title}{Communication-Efficient Learning of Deep Networks
  from Decentralized Data}.
\newblock   (\bibinfo{year}{2016}).
\newblock
\showURL{%
\url{http://arxiv.org/abs/1602.05629}}


\bibitem[\protect\citeauthoryear{Owens, Houston, Luebke, Green, Stone, and
  Phillips}{Owens et~al\mbox{.}}{2008}]%
        {gpu-computing}
\bibfield{author}{\bibinfo{person}{John~D. Owens}, \bibinfo{person}{Mike
  Houston}, \bibinfo{person}{David Luebke}, \bibinfo{person}{Simon Green},
  \bibinfo{person}{John~E. Stone}, {and} \bibinfo{person}{James~C. Phillips}.}
  \bibinfo{year}{2008}\natexlab{}.
\newblock \showarticletitle{GPU Computing}.
\newblock \bibinfo{journal}{{\it Proc. IEEE}} \bibinfo{volume}{96},
  \bibinfo{number}{5} (\bibinfo{date}{May} \bibinfo{year}{2008}),
  \bibinfo{pages}{879--899}.
\newblock


\bibitem[\protect\citeauthoryear{Pan, Chen, Monga, Bengio, and
  J{\'{o}}zefowicz}{Pan et~al\mbox{.}}{2017}]%
        {google-sgd}
\bibfield{author}{\bibinfo{person}{Xinghao Pan}, \bibinfo{person}{Jianmin
  Chen}, \bibinfo{person}{Rajat Monga}, \bibinfo{person}{Samy Bengio}, {and}
  \bibinfo{person}{Rafal J{\'{o}}zefowicz}.} \bibinfo{year}{2017}\natexlab{}.
\newblock \showarticletitle{Revisiting Distributed Synchronous {SGD}}.
\newblock \bibinfo{journal}{{\em CoRR\/}}  \bibinfo{volume}{abs/1702.05800}
  (\bibinfo{year}{2017}).
\newblock
\showURL{%
\url{http://arxiv.org/abs/1702.05800}}


\bibitem[\protect\citeauthoryear{Patarasuk and Yuan}{Patarasuk and
  Yuan}{2009}]%
        {allreduce-yuan}
\bibfield{author}{\bibinfo{person}{Pitch Patarasuk} {and} \bibinfo{person}{Xin
  Yuan}.} \bibinfo{year}{2009}\natexlab{}.
\newblock \showarticletitle{Bandwidth optimal all-reduce algorithms for
  clusters of workstations}.
\newblock \bibinfo{journal}{{\em J. Parallel Distrib. Comput.\/}}
  \bibinfo{volume}{69}, \bibinfo{number}{2} (\bibinfo{year}{2009}),
  \bibinfo{pages}{117--124}.
\newblock
\showDOI{%
\url{https://doi.org/10.1016/j.jpdc.2008.09.002}}


\bibitem[\protect\citeauthoryear{Recht, Re, Wright, and Niu}{Recht
  et~al\mbox{.}}{2011}]%
        {recht2011hogwild}
\bibfield{author}{\bibinfo{person}{Benjamin Recht},
  \bibinfo{person}{Christopher Re}, \bibinfo{person}{Stephen Wright}, {and}
  \bibinfo{person}{Feng Niu}.} \bibinfo{year}{2011}\natexlab{}.
\newblock \showarticletitle{Hogwild: A lock-free approach to parallelizing
  stochastic gradient descent}. In \bibinfo{booktitle}{{\em Advances in Neural
  Information Processing Systems}}. \bibinfo{pages}{693--701}.
\newblock


\bibitem[\protect\citeauthoryear{Rhu, Gimelshein, Clemons, Zulfiqar, and
  Keckler}{Rhu et~al\mbox{.}}{2016}]%
        {DNN-memory}
\bibfield{author}{\bibinfo{person}{Minsoo Rhu}, \bibinfo{person}{Natalia
  Gimelshein}, \bibinfo{person}{Jason Clemons}, \bibinfo{person}{Arslan
  Zulfiqar}, {and} \bibinfo{person}{Stephen~W. Keckler}.}
  \bibinfo{year}{2016}\natexlab{}.
\newblock \showarticletitle{Virtualizing Deep Neural Networks for
  Memory-Efficient Neural Network Design}.
\newblock \bibinfo{journal}{{\em CoRR\/}}  \bibinfo{volume}{abs/1602.08124}
  (\bibinfo{year}{2016}).
\newblock
\showURL{%
\url{http://arxiv.org/abs/1602.08124}}


\bibitem[\protect\citeauthoryear{Ruder}{Ruder}{2016}]%
        {ruder2016overview}
\bibfield{author}{\bibinfo{person}{Sebastian Ruder}.}
  \bibinfo{year}{2016}\natexlab{}.
\newblock \showarticletitle{An overview of gradient descent optimization
  algorithms}.
\newblock \bibinfo{journal}{{\em arXiv preprint arXiv:1609.04747\/}}
  (\bibinfo{year}{2016}).
\newblock


\bibitem[\protect\citeauthoryear{Russakovsky, Deng, Su, Krause, Satheesh, Ma,
  Huang, Karpathy, Khosla, Bernstein, Berg, and Fei-Fei}{Russakovsky
  et~al\mbox{.}}{2015}]%
        {imagenet}
\bibfield{author}{\bibinfo{person}{Olga Russakovsky}, \bibinfo{person}{Jia
  Deng}, \bibinfo{person}{Hao Su}, \bibinfo{person}{Jonathan Krause},
  \bibinfo{person}{Sanjeev Satheesh}, \bibinfo{person}{Sean Ma},
  \bibinfo{person}{Zhiheng Huang}, \bibinfo{person}{Andrej Karpathy},
  \bibinfo{person}{Aditya Khosla}, \bibinfo{person}{Michael Bernstein},
  \bibinfo{person}{Alexander~C. Berg}, {and} \bibinfo{person}{Li Fei-Fei}.}
  \bibinfo{year}{2015}\natexlab{}.
\newblock \showarticletitle{{ImageNet Large Scale Visual Recognition
  Challenge}}.
\newblock \bibinfo{journal}{{\em International Journal of Computer Vision
  (IJCV)\/}} \bibinfo{volume}{115}, \bibinfo{number}{3} (\bibinfo{year}{2015}),
  \bibinfo{pages}{211--252}.
\newblock
\showDOI{%
\url{https://doi.org/10.1007/s11263-015-0816-y}}


\bibitem[\protect\citeauthoryear{Zhang, Choromanska, and LeCun}{Zhang
  et~al\mbox{.}}{2015a}]%
        {zhang2015deep}
\bibfield{author}{\bibinfo{person}{Sixin Zhang}, \bibinfo{person}{Anna~E
  Choromanska}, {and} \bibinfo{person}{Yann LeCun}.}
  \bibinfo{year}{2015}\natexlab{a}.
\newblock \showarticletitle{Deep learning with elastic averaging SGD}. In
  \bibinfo{booktitle}{{\em Advances in Neural Information Processing Systems}}.
  \bibinfo{pages}{685--693}.
\newblock


\bibitem[\protect\citeauthoryear{Zhang, Gupta, Lian, and Liu}{Zhang
  et~al\mbox{.}}{2015b}]%
        {suyog-staleness}
\bibfield{author}{\bibinfo{person}{Wei Zhang}, \bibinfo{person}{Suyog Gupta},
  \bibinfo{person}{Xiangru Lian}, {and} \bibinfo{person}{Ji Liu}.}
  \bibinfo{year}{2015}\natexlab{b}.
\newblock \showarticletitle{Staleness-aware Async-SGD for Distributed Deep
  Learning}.
\newblock \bibinfo{journal}{{\em CoRR\/}}  \bibinfo{volume}{abs/1511.05950}
  (\bibinfo{year}{2015}).
\newblock
\showURL{%
\url{http://arxiv.org/abs/1511.05950}}


\end{thebibliography}

\end{document}